\documentclass[useAMS,usenatbib]{mn2e}
\usepackage{epsfig,natbib}
\usepackage{myaasmacros,amssymb}
\usepackage{graphicx}
\usepackage{times}

\newcommand{\dd}{\mathrm{d}}
\newcommand{\angst}{\mathrm{\AA}}

\def \hi {{\,\rm H\,{\sc i}}}

\def \cm {{\,\rm cm}}

\def \nhi{N_{\mathrm{HI}}}

\def \ncut {N_{\mathrm{cut}}}

\def \neff {N_{\mathrm{eff}}}
\def \dex {\mathrm{dex}}
\def \ldla {l_{\mathrm{DLA}}}
\def \ndla {n_{\mathrm{DLA}}}
\def \fdla {f_{\mathrm{DLA}}}
\def \ffe {\mathcal{F}_{\mathrm{Fe}}}

\def\lsim{\mathrel{\rlap{\lower3pt\hbox{$\sim$}}
    \raise1pt\hbox{$<$}}}                
\def\gsim{\mathrel{\rlap{\lower3pt\hbox{$\sim$}}
    \raise1pt\hbox{$>$}}}                

\newcommand{\confitab}{
\begin{table*}
\caption{Expected value of and $1 -3\sigma$ confidence intervals for
  various quantities. The first five rows specify the fractional
  completeness of optical estimates for the specified quantities. The
  final two rows refer to the expected log mean observable dust
  absorption specified as the reddening $E_{B-V}$ of the background
  quasar in the rest-frame of the DLA (see also Figure
  \protect\ref{fig:dust-red}).}\label{tab:completeness}
\begin{center}
\begin{tabular}{r|cccc}
\hline
Quantity ($Q$) & $E(Q)$ & \multicolumn{3}{c}{Confidence Intervals} \\
&  &  $67\%$ & $95\%$ & $99.7\%$ \\
\hline
$F(l_{\mathrm{DLA}})\,^{(1)}$ & 0.93 &$(0.90,1.00)$ & $(0.81,1.00)$ & $(0.72,1.00)$ \\
$F(\Omega_{\mathrm{DLA}})\,^{(2)}$ & 0.87 &$(0.81,0.97)$ & $(0.70,1.00)$ & $(0.58,1.00)$ \\
$F(\langle Z \rangle)\,^{(3)}$ & 0.75  &$(0.67,1.00)$ & $(0.44,1.00)$ & $(0.26,1.00)$ \\
$F(\langle Z \rangle_{N_{\mathrm{H}}})\,^{(4)}$ & 0.63 &$(0.43,0.82)$ & $(0.32,1.00)$ & $(0.18,1.00)$ \\
$F(\Omega_{Z,\mathrm{DLA}})\,^{(5)}$ & 0.56 &$(0.30,0.75)$ & $(0.22,0.96)$ & $(0.11,1.00)$ \\
\hline
$\log_{10} E_{B-V}$ (optical) & $-2.4$ & $(-2.8,-1.8)$ & $(-3.2,-1.8)$ & $(-4.3,-1.6)$ \\
$\log_{10} E_{B-V}$ (radio) & $-2.1$ & $(-2.5,-1.5)$ & $(-3.5,-1.1)$ & $(-4.5,-0.8)$ \\
\hline
\end{tabular}
\end{center}
\begin{flushleft}
$^{(1)}$The overall completeness of the optical sample, i.e. the ratio
  of the line density estimated from optical samples to the intrinsic
  line density.  \\ 

$^{(2)}$The fractional completeness of optical estimates of the total
  comoving density of \hi~in DLAs.  \\

$^{(3)}$The ratio of the mean metallicity measured in optical samples
  to the intrinsic value.  \\ 

$^{(4)}$The ratio of the mean column density weighted metallicity
measured in optical samples to the intrinsic value. \\

$^{(5)}$The fractional completeness of optical estimates of the total
comoving density of metals in DLAs.
\end{flushleft}
\end{table*}}

\begin{document}

\title[DLA Dust Biasing]{Dust Biasing of Damped Lyman Alpha Systems: a Bayesian Analysis} \author[A. Pontzen,
  M. Pettini]{Andrew Pontzen$^{1}$\thanks{Email:
    apontzen@ast.cam.ac.uk}, Max Pettini$^{1}$
  \\ $^{1}$Institute of Astronomy, Madingley Road, Cambridge CB3 0HA,
  UK}

\date{Accepted 2008 November 03. Received 2008 October 30; in original form 2008 September 01}
\maketitle

\begin{abstract}
If damped Lyman alpha systems (DLAs) contain even modest amounts of
dust, the ultraviolet luminosity of the background quasar can be
severely diminished. When the spectrum is redshifted, this leads to a
bias in optical surveys for DLAs. Previous estimates of the magnitude
of this effect are in some tension; in particular, the distribution of
DLAs in the $(\nhi, Z)$ (i.e. column-density -- metallicity) plane
has led to claims that we may be missing a considerable fraction of
metal rich, high column density DLAs, whereas radio surveys do not
unveil a substantial population of otherwise hidden systems.

Motivated by this tension, we perform a Bayesian parameter estimation
analysis of a simple dust obscuration model. We include radio and
optical observations of DLAs in our overall likelihood analysis and
show that these do not, in fact, constitute conflicting constraints.

Our model gives statistical limits on the biasing effects of dust,
predicting that only $7\%$ of DLAs are missing from optical samples
due to dust obscuration; at $2 \sigma$ confidence, this figure takes a
maximum value of $17 \%$. This contrasts with recent claims that DLA
incidence rates are underestimated by $30 - 50\%$. Optical measures of
the mean metallicities of DLAs are found to underestimate the true
value by just $0.1\,\dex$ (or at most $0.4\, \dex$, $2 \sigma$
confidence limit), in agreement with the radio survey results of
Akerman et al. As an independent test, we use our model to make a
rough prediction for dust reddening of the background quasar. We find
a mean reddening in the DLA rest frame of $\log_{10} \langle E_{B-V}
\rangle \simeq -2.4 \pm 0.6$, consistent with direct analysis of the
SDSS quasar population by Vladilo et al., $\log_{10} \langle E_{B-V} \rangle
=-2.2 \pm 0.1$.  The quantity most affected by dust biasing is the
total cosmic density of metals in DLAs, $\Omega_{Z,\mathrm{DLA}}$,
which is underestimated in optical surveys by a factor of
approximately two.
\end{abstract}


\section{Introduction}

\begin{figure*}
\includegraphics[width=1.0\textwidth]{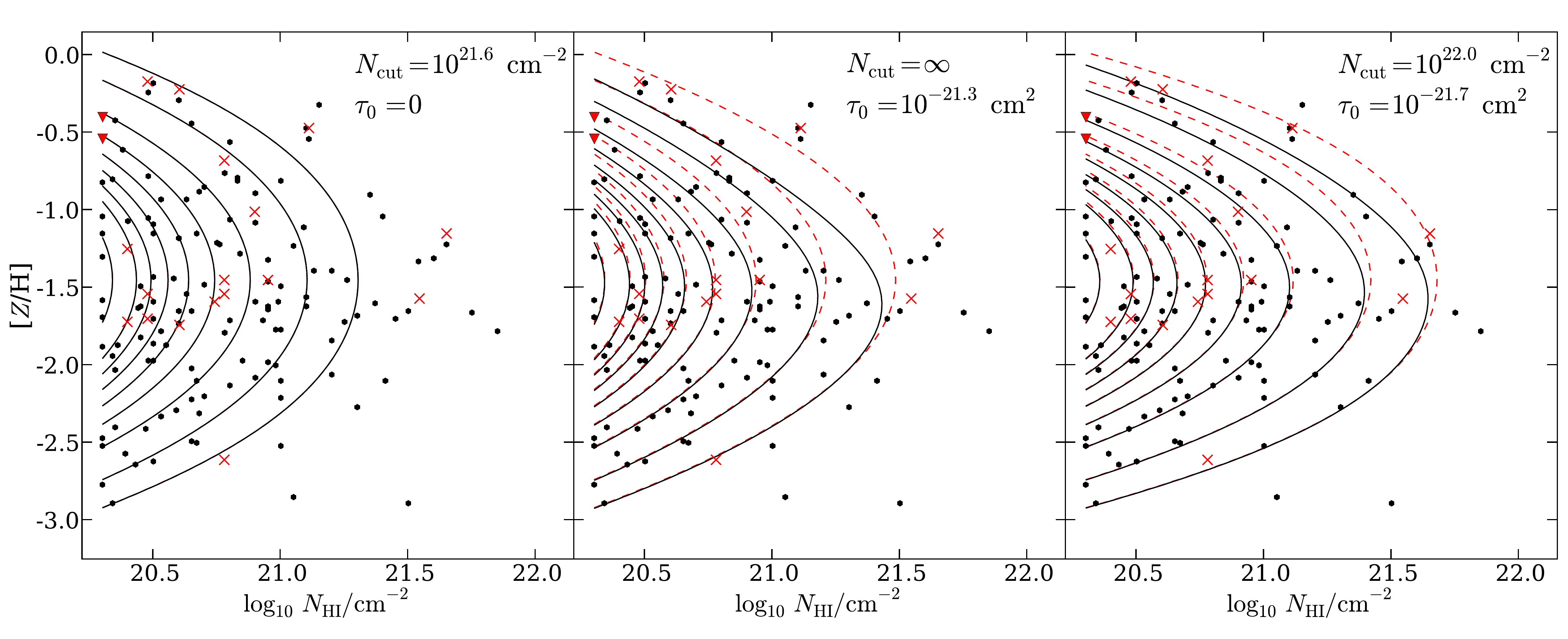}
\caption{A heuristic picture of different regions of our model
  space. Our optical sample based on \protect\cite{DZinprep} is shown
  by dots; the \textsc{Corals} radio sample based on
  \protect\cite{2005A&A...440..499A} is shown by crosses or triangles
  for the upper limits (see Section \ref{sec:likelihood} for a
  discussion of these datasets). Contour lines of equal probability
  density in the $\log\,N_{\mathrm{HI}}$ -- $\log\,Z$ plane for
  finding a DLA in an optical sample (solid lines) or radio sample
  (dashed lines where this differs from the optical case) are shown
  for three models. In each panel, the contours correspond, from left
  to right, to $0.9,0.8,\dots,0.2,0.1$ and $0.05$ times the peak
  probability density.  In the left panel, high column densities of
  hydrogen are intrinsically unlikely but there is no dust absorption;
  in the central panel, the Schechter-type intrinsic cut-off is
  absent, but dust truncates the observed high column densities of
  metals. The final panel illustrates the favoured type of model from
  our analysis, in which both intrinsic and obscuration mechanisms
  have some part to play in shaping the optically observed
  distribution. By eye, the three sets of optical contours appear
  similar; this illustrates the need for a rigorous method to probe
  the role of dust in shaping the final
  distribution.}\label{fig:models}

  
\end{figure*}

Damped Lyman alpha systems (DLAs), neutral gas with column densities
$N_{\mathrm{HI}}> 2 \times 10^{20} \cm^{-2}$ seen in absorption
against more distant luminous sources (generally quasars), are of
substantial interest to observational and computational cosmologists.
Despite disagreement over their precise nature, they are certain to
trace a set of objects which constrain our theories of galaxy
formation. This is guaranteed by the simple observational fact that
they contain the overwhelming majority of neutral hydrogen (a
necessary precursor to molecular hydrogen and therefore star
formation) over all redshifts $z>0$ \citep{1987ApJ...321...49T}. For a
review of observational and theoretical results see
\cite{2005ARA&A..43..861W}.

One area of controversy in the interpretation of DLA observations is
the extent to which biases are introduced by dust: it is possible to
imagine scenarios in which certain metal rich, high column density
DLAs dim their background quasars such that significant fractions are
not detected in optical surveys.  Early attempts at assessing the
magnitude of this effect by comparing the spectral slopes of QSOs with
and without intervening DLAs seemed to suggest that estimates of
important quantities such as the total density of neutral hydrogen
($\Omega_{\mathrm{DLA}}$) and the mean metallicity ($\langle Z
\rangle$) of DLAs could be incorrect by orders of magnitude
\citep[][and references therein]{1993ApJ...402..479F}. While recent
results
\cite[e.g.][]{2004MNRAS.354L..31M,2005AJ....130.1345E,2008A&A...478..701V}
show that the extent of dust reddening was substantially overestimated
in these early works, emphasis on the observational evidence in
apparent support of the obscuration scenario has shifted to the
distribution of absorbers in $(\nhi,Z)$ space. First noted by
\cite{1998A&A...333..841B}, there is a dearth of absorbers exhibiting
simultaneously high $\nhi$ {\it and} high $Z$ -- exactly as would be
expected in a scenario invoking significant dust absorption.  Recent
work on such models \citep[e.g.][]{2005A&A...444..461V}
has suggested that smaller but still important effects arise from dust
obscuration.  In particular, a dust-induced bias has been invoked by
simulators to reconcile high DLA metallicities encountered in models
with the generally low values measured empirically
\citep{2003ApJ...598..741C,2004MNRAS.348..435N}.

One should be clear, however, that this interpretation is not unique
-- when at the tails of both $\nhi$ and $Z$ distributions, systems
will anyway be rare (see Figure \ref{fig:models}, which shows similar
distributions arising from the different models described in Section
\ref{sec:model}). Thus statistical interpretation of these apparent
trends must be approached with care. 

In fact, a range of constraints cast doubt on models predicting a
substantial bias. Starting from the observed relative abundances of
elements which are depleted onto dust by differing amounts such as
zinc (undepleted) and chromium (severely depleted),
\cite{1997ApJ...478..536P} estimated a DLA-induced extinction at $1500
\angst$ of just $\sim 0.1\,\mathrm{mag}$.  More directly, samples of
radio-selected QSOs (which are unaffected by dust) exhibit similar
incidence rates of DLAs as their optical counterparts
(\citeauthor{2001A&A...379..393E} \citeyear{2001A&A...379..393E}; see
also \citeauthor{2006ApJ...646..730J} \citeyear{2006ApJ...646..730J},
although the optical identification in this latter work is not
complete). Moreover, high resolution spectroscopy of the
\citeauthor{2001A&A...379..393E} DLAs (known as the \textsc{Corals}
sample) shows a similar distribution of metal column densities as
found in optical samples \citep{2005A&A...440..499A}. While the
radio-selected samples do show a marginal $1\sigma$ difference from
the optical data in both mean metallicity and incidence rate of DLAs
(in the correct sense for a dust obscuration bias signature), it is
not at all clear whether this is merely a statistical fluke.  It is
worth noting that, even if the effect on measures such as the
incidence rate and mean metallicity is minor, weighted measures such
as $\Omega_{\mathrm{DLA}}$ can be more critically affected. At most
risk of being underestimated is the total mass of metals in DLAs,
$\Omega_{\mathrm{Z,DLA}}$, which is observationally interesting when
conducting a census of metal enrichment over cosmic time \citep[][and
  references therein]{PetMet,2007MNRAS.378..525B}.

Overall, the previous work described above appears to be in some
tension. Difficulty in understanding these tensions is exacerbated by
analyses using ad-hoc statistical methods or ``by-eye''
assertions. These problems motivated the present work in which we have
taken a Bayesian parameter estimation approach to putting useful
limits on the effects in question. In our analysis we have used four
logically distinct observational datasets: an optically selected
sample of DLAs, a radio selected sample of DLAs, SDSS\footnote{Sloan
  Digital Sky Survey} statistics for the column densities of DLAs, and
overall incidence rates for DLAs in radio and optical surveys.

The Bayesian parameter estimation formalism requires us to (i)
formulate a parameterized model describing the data and (ii) place
prior probabilities on the distribution of parameters for the model.
These two processes can, of course, give rise to controversy~--
especially when the physical processes in play are hard to model. In
particular, stage (i) places a unit prior probability on our chosen
model: we might humbly admit that this is not entirely satisfactory,
but emphasize that the Bayesian technique does not introduce but
merely highlights such difficulties. We also performed additional
analysis on a widened parameter space which goes some way to
mitigating our concerns (see Appendix \ref{sec:choice-fit-funct}). For
more details on the Bayesian technique see
e.g. \cite{2003prth.book.....J}.

The remainder of this paper is structured as follows. In section
\ref{sec:model}, we develop a basic model which we argue captures the
significant effects of dust-induced obscuration. We describe our use
of optical- and radio-selected survey results to calculate likelihoods
for this model in section \ref{sec:likelihood}. With some simple
priors described in section \ref{sec:priors}, we examine the resulting
statistical estimates for completeness of optical samples in section
\ref{sec:results}. Finally, we conclude that dust biasing is a real
but minor effect in section \ref{sec:conclusions} in which we also
discuss how our technique and results differ from similar work by
\cite{2005A&A...444..461V}.

\section{Model and Parameter Estimation}\label{sec:model--parameter}

In this section, we will form a simple model for the observed
behaviour of absorbers with a continuous parameter which describes the
extent to which dust obscuration plays a part. Performing parameter
estimation will then allow us to assess the effect of dust
absorption on the observed statistics. The final model has five
parameters, so we use a Metropolis-Hastings Markov chain Monte Carlo
algorithm to sample the posterior probability distribution
\citep[][and references therein]{1992nrca.book.....P}.

\subsection{Model}\label{sec:model}

Our simple model starts from the assumption that the intrinsic
distribution of DLAs is separable in the $(\nhi, Z)$ plane.  Although
locally $N_{\mathrm{HI}}$ and the star formation rate may be expected
to be correlated (via the Schmidt-Kennicutt relation observed in local
galaxies, see e.g. \citeauthor{1998ApJ...498..541K}
\citeyear{1998ApJ...498..541K}), our own $N$-body simulations of
galaxy formation \citep{PontzenDLA}, as well as previous simulations
\citep{2003ApJ...598..741C,2004MNRAS.348..435N}, suggest that there is
no significant correlation between $N_{\mathrm{HI}}$ and the global
star formation history of the host galaxy and hence its
metallicity. Thus
\begin{equation}
f_{\mathrm{DLA}}(N_{\mathrm{HI}},Z) = f_N(N_{\mathrm{HI}}) \, f_Z(Z) \label{eq:intrinsic-distrib}
\end{equation}
where $f_{\mathrm{DLA}}(N_{\mathrm{HI}},Z)$ gives the intrinsic
probability density of a DLA's location in the $(\nhi,Z)$ plane,
picked with no observational biases. The distribution of column
densities $f_N$ follows a Schechter function \citep[as suggested
  by][]{1995ApJ...454...69P}
\begin{equation}
 f_{\mathrm{N}}(N_{\mathrm{HI}})  =  N_{\mathrm{HI}}^{\alpha} \, e^{-N_{\mathrm{HI}}/\ncut} \label{eq:colden-schechter}
\end{equation}
where $\alpha$ measures the low column density slope and $\ncut$ is a
characteristic cut-off column density. The distribution of
metallicities $f_Z$ is assumed lognormal
\begin{equation}
f_{Z}(Z)   =  \frac{1}{Z} \exp \left(-\frac{([Z] - \mu_Z)^2}{2 \sigma_Z^2}\right)\label{eq:metal-lognormal}
\end{equation}
where $[Z]=\log_{10} Z/Z_{\odot}$, $\mu_Z$ is the mean log metallicity
and $\sigma_Z$ is the standard deviation of the log metallicity. We
have intentionally not normalized our distribution functions at this
stage.

These functional forms are based jointly on observational and
simulated work -- but of course the observations are from magnitude
limited optical samples, so it is worth asking whether the intrinsic
distributions could in fact have a substantially different shape; we
have addressed this possibility in Appendix A (but find that our
current parameterization is adequate given some fairly weak
assumptions).

\subsubsection{Determination of dust column density}

\begin{figure}
\includegraphics[width=0.5\textwidth]{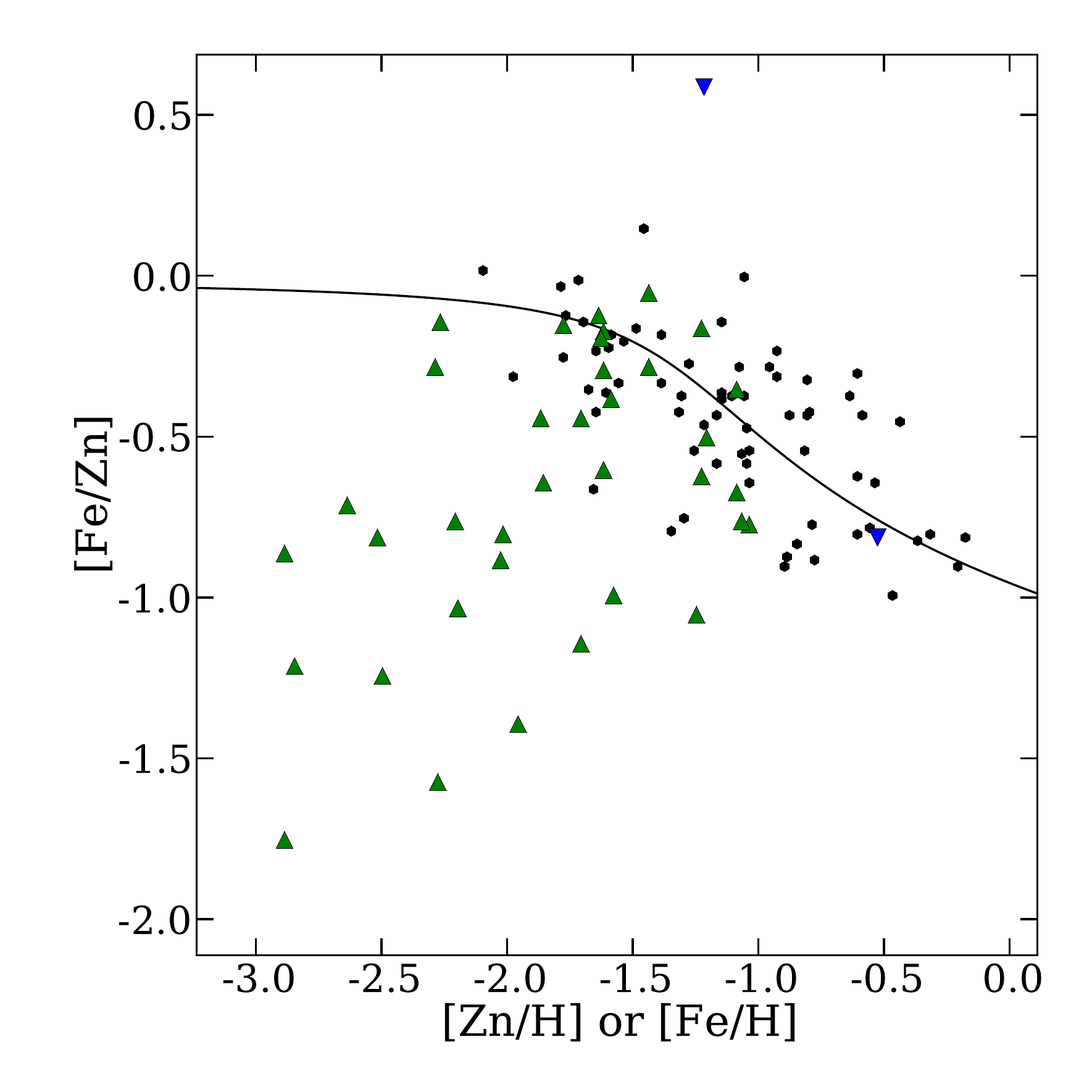}
\caption{The adopted relation
  \protect\cite[from][]{2005A&A...444..461V} between the metallicity
  and fraction of iron in dust is based on the observed relation
  between metallicity and iron-to-zinc ratio (since zinc does not
  deplete onto dust grains). Here, data from the optical sample
  described in the Section \ref{sec:likelihood} are plotted (dots and
  triangles for limits) along with the best fit model
  (curve). }\label{fig:vp05-depletion}
\end{figure}

We will assume that the optical depth of dust in any system may be modelled
as
\begin{equation}
\tau_{\mathrm{dust}}(\lambda,\nhi,Z) = \tau_{0}(\lambda)
N_{\mathrm{HI}}
\frac{Z \ffe(Z)}{Z_{\mathrm{0}}\ffe(Z_{0})}\label{eq:dust-scaling}
\end{equation}
where $Z_0$ is a normalization metallicity, $\ffe$ represents the
varying fraction of iron in the dust phase as a function of
metallicity, and $\tau_0(\lambda)$ specifies a linear scaling between
dust column density and optical depth at wavelength $\lambda$.  This
form is based on the fair assumption that the DLA gas density is
dominated by the neutral hydrogen density
\cite[see][]{2005ARA&A..43..861W}.  Our results are actually rather
insensitive to the exact functional form of $\ffe$, so long as the
fraction in dust increases gently with metallicity, but for ease of
comparison we have adopted the form suggested by
\cite{2005A&A...444..461V}:
\begin{equation}
\ffe = \frac{1}{2} + \frac{1}{\pi} \tan^{-1}
\left(\frac{[Z]-[Z]_0} {\Delta
  [Z]}\right)\textrm{.}\label{eq:vp05-dfe}
\end{equation}
 Since zinc does not deplete onto dust, the ratio of iron to zinc
 column densities is predicted by $\ffe$:
\begin{equation}
\mathrm{[Fe/Zn]} \equiv
\log_{10}\left(1-\ffe\right)\label{eq:fFe-to-ratio}
\end{equation}
This relation can be used, along with observational constraints on
iron and zinc abundances in individual systems from the dataset
described below in Section \ref{sec:likelihood}, to estimate best fit
parameters\footnote{We could have included this estimation in our full
  Bayesian formalism, but due to the insensitivity of our results to
  the details of the relation $\ffe(Z)$, such an approach would add
  complexity without substantial benefit.}  $[Z]_0=-1.3$ and $\Delta[Z]=0.48$
in equation (\ref{eq:vp05-dfe}). For reference, we have plotted this
relationship in Figure \ref{fig:vp05-depletion}.

\subsubsection{Conversion from optical depth to detection probability}
We base our predictions for optical samples on the behaviour of a
survey for quasars in the SDSS $i$ band with a mean wavelength
$\lambda_i\simeq 7480 \angst$. For our optical data (defined in
section \ref{sec:likelihood} below), the mean redshift $\langle
z_{\mathrm{DLA}} \rangle=3.0$ translates into a DLA rest-frame
wavelength of $\lambda_0 = \lambda_i / (1+\langle z_{\mathrm{DLA}} \rangle)\simeq
1900 \angst$. This will be useful in fixing a prior on $\tau_0$ later.
\begin{figure}
\includegraphics[width=0.5\textwidth]{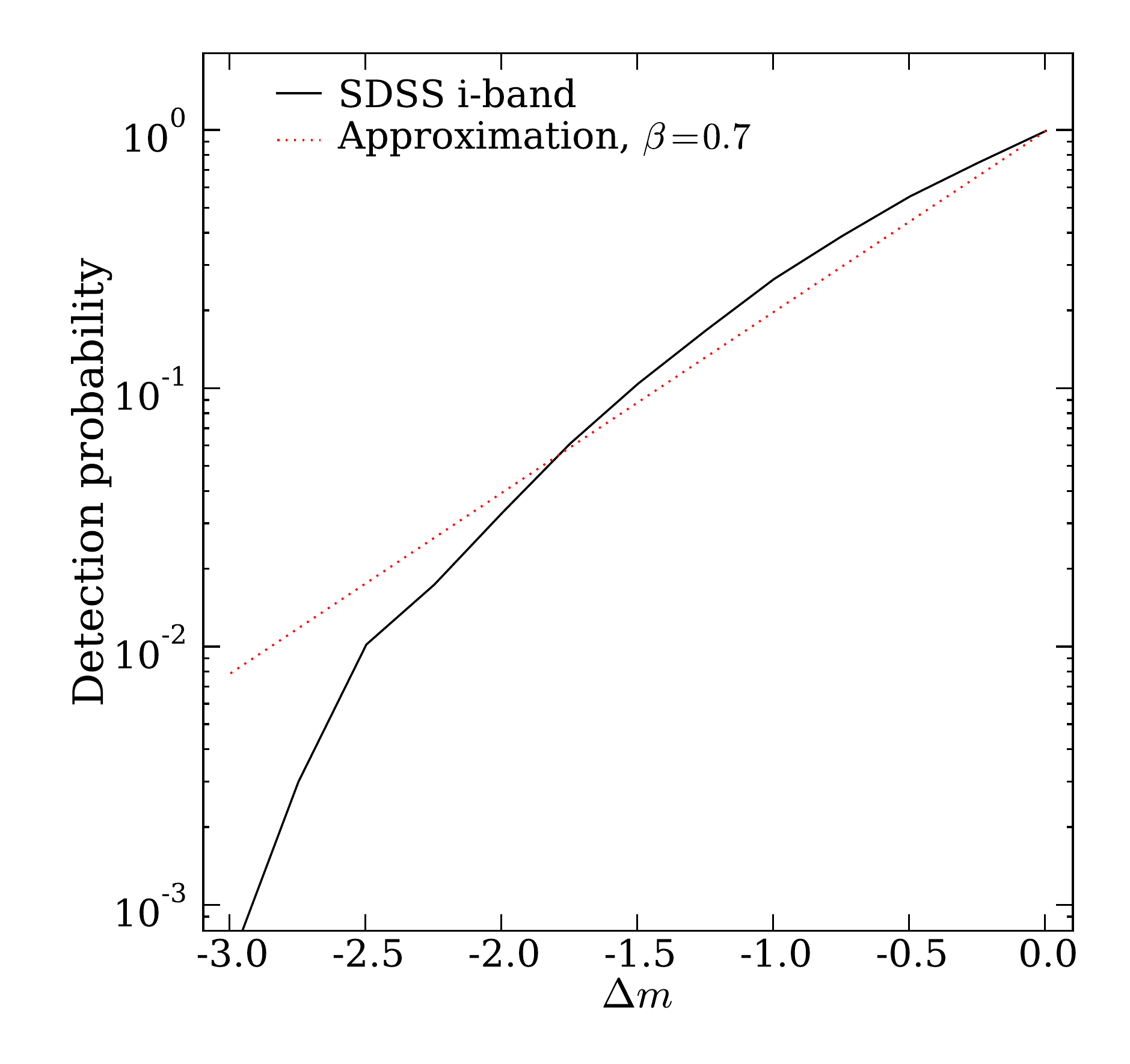}
\caption{The completeness function for SDSS QSOs in the $i$ band
  (solid line) compared to our simple analytic approach (dotted
  line). $\Delta m$ is the change in the apparent $i$-magnitude, while
  the vertical axis shows the fraction of QSOs that would be missed
  under such a reduction in brightness assuming a step-function
  sensitivity. The differences between the exact and simple analytic
  model are only significant when the probability of detection falls
  below $10^{-2}$, and thus have little impact on our overall
  statistics.  }\label{fig:sdss-obs-prob}
\end{figure}
Over the full range of our sample ($1.8<z<3.5$) the DLA rest-frame
wavelength varies between $1600 < \lambda_0/\angst < 2700$.  According
to the low-metallicity extinction law measured in the SMC \citep[Small
  Magellanic Cloud;][]{1992ApJ...395..130P}, there is a factor two
variation in the expected strength of dust absorption over the
interval $1600 < \lambda_0/\angst < 2700$. However, we have chosen not
to implement the resulting redshift dependences since:
\begin{enumerate}
\item the optical metallicity data (section \ref{sec:likelihood}) are
  not compiled from a single magnitude-limited sample, but rather from
  multiple datasets so that an exact modelling is for all practical
  purposes impossible;
\item computationally, it would be extremely expensive to allow
  redshift variation (since the normalization of the models would need
  to be recalculated for every absorber, instead of once per model -- see
  equations~\ref{eq:n0},~\ref{eq:f0} below);
\item the stochastic variation in other parameters (metallicity and
  column density) dwarfs the maximum variation of a factor two; and
\item since the metallicity and column density evolution is known to
  be weak \citep{2005ARA&A..43..861W}, systematic biases are unlikely
  to arise from neglecting slight redshift dependences.
\end{enumerate}

For similar reasons, we do not allow ourselves to become fixated on an
exact modelling of the observed quasar background luminosity function.
We assume the background population of quasars has an observed
distribution in the SDSS $i$ band which obeys $\log_{10} \dd N/ \dd m
= c + \beta m$ with $\beta \simeq 0.7$ \citep{2006AJ....131.2766R}
and that the detection probability is one above a given brightness
threshold ($m<m_0$) and zero below ($m>m_0$). Then the total number of
quasars which can be observed in the absence of dust obscuration is
\begin{equation}
N_{\mathrm{tot,unobsc}} = \int_{-\infty}^{m_0} 10^{c + \beta m} \dd m \textrm{.}
\end{equation}
When dust is introduced, the luminosity of a given system is reduced
by a factor $e^{-\tau}$; the apparent magnitude changes by $\Delta
m(\tau) = 2.5 \tau / \ln 10$. The number of quasars which could be
observed in the presence of such absorption is
\begin{equation}
N_{\mathrm{tot,obsc}}(\tau) = \int_{-\infty}^{m_0} 10^{c + \beta[m+\Delta m(\tau)]} \dd m 
\end{equation}
so that the probability of detecting a system which has optical
depth $\tau$ in the $i$ band is reduced by the factor 
\begin{equation}
p_{\mathrm{detect}}(\tau) = \frac{N_{\mathrm{tot,obsc}}(\tau)}{N_{\mathrm{tot,unobsc}}} = \exp(-2.5 \beta \tau) \textrm{.}\label{eq:p-det}
\end{equation}
 While emphasis has previously been placed on departures of $\dd N /
 \dd m$ from power law behaviour and hence more complex forms of
 $p_{\mathrm{detect}}$ \citep{2004ApJ...615..118E}, we find that the
 actual effects are well modelled by our approach (see Figure
 \ref{fig:sdss-obs-prob} which compares the analytic and exact SDSS
 models). Other than simplicity for its own sake, there is a tangible
 benefit to keeping the model basic: it can be integrated partially
 analytically (see Section \ref{sec:normalization} below).

\begin{table*}
\caption{Summary of parameters and priors thereon for the DLA observation model.}\label{tab:priors}
\begin{tabular}{lcll}
Parameter & Equation & Description & Prior \\ 
\hline 
$\alpha$ & (\ref{eq:colden-schechter}) &  Low $N_{\mathrm{HI}}$ slope &  Flat $-2.5<\alpha<0$ \\ 
$\ncut$ & (\ref{eq:colden-schechter}) & Intrinsic $N_{\mathrm{HI}}$ rolloff scale & Flat in log space;
$\ncut<10^{24}\, \cm^{-2}$ \\
$\mu_Z$, $\sigma_Z$ & (\ref{eq:metal-lognormal}) & Parameters for lognormal metallicity
distribution & Flat priors $-3<\mu_Z<0$; $0.1<\sigma_Z<3.0$ \\
$ \tau_0$ & (\ref{eq:dust-scaling})  & Dust optical depth normalization for SMC ($Z_0 =
Z_{\odot}/6$) & $\mu=-21.7$, $\sigma = 1$ \\

\hline
\end{tabular}

\end{table*}

\subsubsection{Completing the model}

The optically observed joint probability distribution
$n_{\mathrm{DLA}}(\nhi,Z)$ is simply the product of the intrinsic
$f_{\mathrm{DLA}}$ with the detection probability
$p_{\mathrm{detect}}$,
\begin{eqnarray}
  n_{\mathrm{DLA}}(N_{\mathrm{HI}},Z) & = & 
  f_{\mathrm{DLA}}(N_{\mathrm{HI}},Z) \nonumber \\
 &&\hspace{-1mm}\times\exp
  \left\{-2.5 \beta  \tau_0 \nhi \frac{Z\,\ffe(Z)}{Z_0 \ffe(Z_0)} \right\}\label{eq:optical-distrib}
\end{eqnarray}
which completes our model. The five free parameters that we have
introduced are summarised in Table~\ref{tab:priors}.

In Figure \ref{fig:models} we have illustrated some distributions
which can be achieved by the model above. The dotted and solid
contours trace respectively lines of constant $\fdla$ (the intrinsic
distribution, equation \ref{eq:intrinsic-distrib}) and $\ndla$ (which
includes the effect of dust obscuration, equation
\ref{eq:optical-distrib}) -- i.e. the former traces the distribution
of radio-selected DLAs and the latter that of optically-selected
DLAs. We have also plotted our optical sample (dots) and radio sample
(crosses or triangles for upper limits) on each panel -- see Section
\ref{sec:likelihood} below for details of these datasets. The left
panel shows a model where $\tau_0=0$ (so no dust obscuration effects
are in play). One should note that, despite this, the contours show
very small probabilities for high $\nhi$, high $Z$ absorbers simply
because the underlying separable distribution $f_N f_Z$ predicts few
absorbers in this region. The second panel shows a model similar to
that used by \cite{2005A&A...444..461V}, where no intrinsic cut-off occurs at high $\nhi$,
but dust obscuration hides the higher column densities.  The third
panel shows a combination of these two effects forming the final
distribution; our results (Section \ref{sec:results}, Figure
\ref{fig:main-likelihoods}) will show that such a combination is
necessary to best describe the data.

\subsubsection{Normalization}\label{sec:normalization}

In assessing our likelihood, we will split the data into two
logically distinct constraints: the total density of systems 
and the distribution of systems within the $(\nhi,Z)$ plane.
For these purposes, we require the normalizing constants
\begin{eqnarray}
n_0 & = &  \int \dd \nhi\, \dd Z\, n_{\mathrm{DLA}}(\nhi,Z) \label{eq:n0} \\
f_0 & = &  \int \dd \nhi\, \dd Z\, f_{\mathrm{DLA}}(\nhi,Z) \textrm{ .}\label{eq:f0}
\end{eqnarray}
Because of the separability of the intrinsic absorption $f$, its
normalizing constant $f_0$ may be calculated straight-forwardly to be
\begin{equation}
f_0 = \sqrt{2 \pi \sigma_Z \ln 10} \, \ncut^{1+\alpha}\,  \Gamma(1+\alpha,N_0/\ncut) 
\end{equation}
where $N_0=2 \times 10^{20} \cm^{-2}$ is the DLA limiting column
density and $\Gamma$ is the incomplete gamma function,
\begin{equation}
\Gamma(a,x) \equiv \int_x^{\infty} \dd t\, t^{a-1} e^{-t}
\end{equation}
for the evaluation of which we employ a standard numerical algorithm
\citep{1992nrca.book.....P}.  For the obscured case we integrate
analytically over $\nhi$, but the metallicity integral must be
performed numerically:
\begin{equation}
n_0 = \int_{0}^{\infty} \dd Z \, f_Z(Z)\, \neff^{1+\alpha}\,  \Gamma(1+\alpha,N_0/\neff(Z))
\end{equation}
\begin{equation}
\textrm{where }\neff(Z)^{-1} = \ncut^{-1} + 2.5 \beta \tau_0
\frac{\ffe(Z)Z}{\ffe(Z_0)Z_0}
\end{equation}
and $f_Z(Z)$ is defined by equation~(\ref{eq:metal-lognormal}).

\subsection{Data and Likelihood}\label{sec:likelihood}

Each model is assessed on four points corresponding to properties of
optically selected absorbers, properties of radio selected absorbers,
a comparison of the line densities of absorbers in these two types of
survey and finally SDSS constraints on the column density
distribution. The overall likelihood $\mathcal{L}$ is simply the
product of the four factors:
\begin{equation}
\mathcal{L} = \mathcal{L}_{\mathrm{opt}} \, \mathcal{L}_{\mathrm{rad}}\,  \mathcal{L}_{\mathrm{linedens}}\,  \mathcal{L}_{\mathrm{SDSS}} \label{eq:tot-likeli}
\end{equation}
with the terms formally defined below in equations
(\ref{eq:opt-likeli} -- \ref{eq:sdss-likeli}).

We use data from high-resolution optical measurements of 123 DLAs
based on the compilation by \cite{DZinprep} restricted to the redshift
range $1.8<z<3.5$ to match the approximate range of the
\textsc{Corals} radio sample (see below)\footnote{We also checked that
  we obtained compatible, although slightly less well constrained,
  final results from the smaller optical sample described by
  \cite{2007ApJS..171...29P}.}. For each DLA we use as a measure of
its metallicity the zinc abundance relative to the preferred solar
value $12 + \log_{10}\,\left(\mathrm{Zn/H}\right)_{\odot} = 4.63$
  \cite[from][]{2003ApJ...591.1220L}; where zinc measurements are
  unavailable we use the iron abundance normalized similarly by
  $12 + \log_{10}\,\left(\mathrm{Fe/H}\right)_{\odot} = 7.47$ . Zinc
  is not prone to deplete onto dust grains but at lower column
  densities its transitions become too weak for measurement;
  conversely, iron is typically disfavoured as a metallicity indicator
  since it is refractory but the depletion is small at low
  metallicities (see Figure~\ref{fig:vp05-depletion}, in which the
  depletion of iron relative to zinc is plotted).  Since our sample is
  dominated by zinc measurements down to $Z \sim Z_{\odot}/30$, the
  iron depletion should not be a major concern.  In any case, any
  systematic underestimates of metallicities which may arise would
  apply equally to radio- and optically-selected DLAs and should
  therefore not result in any substantive systematic biases for our
  test.

Observers typically favour targeting high $N_{\mathrm{HI}}$ systems
for high resolution follow-up. For this reason, we do not allow the
distribution of $\nhi$ values in the optical metallicity sample to
affect our statistics, instead restricting ourselves to measuring the
likelihood of each metallicity observation with the column density of
the responsible absorber as a given, i.e.
\begin{equation}
 \mathcal{L}_{\mathrm{opt}} = \prod_i p_{\mathrm{opt}}(Z_i|N_i) = \prod_i \frac{n(N_i,Z_i)}{ \int_0^{\infty} n(N_i,Z) \dd Z}\label{eq:opt-likeli}
\end{equation}
where $i$ ranges over the optical sample and the final relation
follows from the conditional probability rule: $p(Z_i|N_i)\, p(N_i) =
p(Z_i \, \& \,N_i)$.  

Metallicity data from the radio-selected {\textsc{Corals}} survey are
taken from \cite{2005A&A...440..499A}. As for the optical sample
described above, we use Zn or (where necessary) Fe to define the
metallicity. Unlike the optical case, each radio observation is
assessed jointly on its column density and metallicity since no column
density biases are expected.  For two DLAs no abundances have been
measured and for a further two only an upper limit on the metallicity
is available. By noting that this situation corresponds to an
``infinite upper limit'' on the metallicity of the former two DLAs, we
can include all systems consistently in the likelihood:
\begin{equation}
 \mathcal{L}_{\mathrm{rad}} = \prod_i \frac{f(N_i,Z_i)}{f_0} \times \prod_j \frac{1}{f_0} \int_{0}^{Z_{\mathrm{max},j}} f(N_j,Z) \dd Z
\end{equation}
where $i$ ranges over the radio sample with measured metallicities,
and $j$ ranges over those four with only upper limits.

Separately from the $(\nhi, Z)$ distribution of observed DLAs, we
should also consider the overall incidence rate in optical (e.g. SDSS)
and radio surveys. The incidence of DLAs in the SDSS has been
discussed extensively for the third data release (DR3) by
\cite{2005ApJ...635..123P}; here we will make use of the updated
statistics for DR5\footnote{{\tt
    www.ucolick.org/\~{}xavier/SDSSDLA/DR5/}}.

Because the SDSS pathlength is very much larger than that of any other
survey, we make the simplifying assumption that there is no error on
its determination of the obscured rate of DLA incidence,
$l_{\mathrm{obsc}}=0.063$ over $2.2<z<3.5$. This quantity is a
measurement of the number density of DLAs per unit ``absorption
distance'' $X$, defined by $\dd X/\dd z= H_0(1+z)^2/H(z)$.  The line
density of radio-selected quasar DLAs is increased by the ratio of all
DLAs to unobscured, i.e. $l_{\mathrm{unobsc}} = l_{\mathrm{obsc}}
f_0/n_0$ (see equations \ref{eq:n0}, \ref{eq:f0}). This follows
because the line density is at first order proportional to the
normalizing constants $f_0$ and $n_0$ for the unobscured and obscured
cases respectively (see also discussion around equation
\ref{eq:l-dla}).

We make use of the \textsc{Corals} \citep{2001A&A...379..393E} results
giving a radio sample pathlength of $\Delta X = 195$ (assuming
$\Omega_M=0.3$, $\Omega_{\Lambda}=0.7$ whence $\dd X/\dd z\simeq 3.5$,
any errors in which are small compared to sample variance). Although
\cite{2006ApJ...646..730J} present additional radio-selected DLA
statistics, unlike the \textsc{Corals} results their optical
identifications are incomplete and so to be conservative we did not
take advantage of the expanded sample.  The overall expected number of
DLAs in the \textsc{Corals} sample is $\lambda$ where
\begin{equation}
\lambda = l_{\mathrm{unobsc}} \Delta X = l_{\mathrm{obsc}} \Delta X \frac{f_0}{n_0} = 12.3 \frac{f_0}{n_0} \textrm{ .}
\end{equation}
For each model $f_0/n_0$ and hence $\lambda$ is determined; given the
fixed number of DLAs actually seen in the sample, $k=17$, the
corresponding likelihood is given by the Poisson distribution
\begin{equation}
 \mathcal{L}_{\mathrm{linedens}} = \frac{\lambda^k e^{-\lambda}}{k!} \textrm{ .}\label{eq:linedens-likeli}
\end{equation}
We note that the mean redshift $\langle z \rangle$ of all DLAs in the
two samples (radio and $z$-limited SDSS) is respectively 2.5 and
2.9. Given the very slow evolution of DLA incidence rate at high
redshift, this difference is unimportant.

Finally, the SDSS data produce a joint constraint on the strength of
dust absorption $\tau_0$ and the Schechter function cut-off $\ncut$
through the distribution of column densities. Because our optical data
likelihood $\mathcal{L}_{\mathrm{opt}}$ does not take account of the
distribution of column densities, the SDSS survey may be regarded as
an entirely independent constraint with likelihood
\begin{equation}
 \mathcal{L}_{\mathrm{SDSS}} =  \prod_i \frac{1}{n_0} \int_0^{\infty} n(N_i,Z) \dd Z \label{eq:sdss-likeli}
\end{equation}
where $i$ ranges over the 587 systems in the previously described
subset of the SDSS DR5 data.

For readers unused to the Bayesian approach to statistics, it may be a
surprise that our likelihoods are a product of probability {\it
  densities} and therefore will vary under reparameterizations of the
data. However the final analysis considers only ratios of
probabilities for different models, for which the Jacobian factors
cancel.

\subsection{Priors}\label{sec:priors}

As discussed in the introduction, we are required to place prior
probability distribution functions on our parameters to summarise
known physics and observational constraints not included in the
likelihood. We have little information on DLA dust absorption which is
not used in our likelihood analysis, so most of our priors are
deliberately as neutral as possible, while limiting the values to
reasonable physical expectations.

Note that allowing the column density distribution function cut-off to
tend to infinity ($\ncut \to \infty$) allows for significant numbers
of implausibly dense environments. A conservative prior from
observations of extreme astrophysical situations (in particular,
active galactic nuclei and gamma ray bursts) is that column densities
do not exceed $N_{\mathrm{H}} \sim 10^{24} \cm^{-2}$, which is
implemented by adopting a flat log prior for $\ncut<10^{24}
\cm^{-2}$. In practice, the likelihood is sharply peaked around $\ncut
\sim 10^{21.6} \,\cm^{-2}$ so that our results are insensitive to this
choice.

The parameter controlling the strength of the dust extinction effect,
$\tau_0$, can be estimated. We use the SMC extinction curve
\citep{1992ApJ...395..130P} at our mean rest-frame wavelength
$\lambda_0\simeq 1900 \angst$ (see section \ref{sec:model} above),
gaining $\tau(\lambda_0) = 10^{-21.7} \, (\nhi/\cm^{-2})$. If we were
confident of this estimate, we could fix $\tau_0=10^{-21.7} \,\cm^2$
and $Z_0=Z_{\odot}/6$ (the SMC metallicity) in equation
(\ref{eq:dust-scaling}). However, when searching for {\it direct}
evidence of dust obscuration, this would appear circular: $\tau_0$
should be allowed to vary. $Z_0$ does not need to vary, even if we are
unsure of the exact SMC metallicity, since a misestimate can simply be
absorbed into the posterior value of $\tau_0$ without affecting the
observable predictions of the model.

With our caveat of circularity in mind, it is tempting to try and
place some form of uniform prior on $\ln \tau_0$ -- but this is
impossible, since as the effect tends to zero ($\ln \tau_0 \to
-\infty$), the models become indistinguishable in their predictions
for a finite data-set and the likelihood density becomes constant. One
must therefore be careful to assign a prior with finite integral as
$\ln \tau_0 \to -\infty$ but which is not so sharp as to exclude the
possibility of an unexpected result\footnote{Such an unexpected result
  would likely point to a deficiency in the model.}. This will anyway
reflect substantial uncertainties in our estimate of $\tau_0$ (and
$Z_0$).  Thus we assign a generous order of magnitude uncertainty at
the $1\sigma$ level, making the prior on $\log_{10} \tau_0/\cm^{2}$
normal with mean $\mu = -21.7$ and variance $\sigma = 1.0\, \dex$. The
effect of this prior on the results is discussed in more detail in
Section \ref{sec:results} below.

We have assigned flat priors to the remaining parameters which control
the intrinsic metallicity model and the weak end of the column density
distribution (see Table \ref{tab:completeness}). These are well
constrained by the data; consequently the priors do not impact strongly 
on our final results.

\section{Results}\label{sec:results}

\begin{figure}
\includegraphics[width=0.5\textwidth]{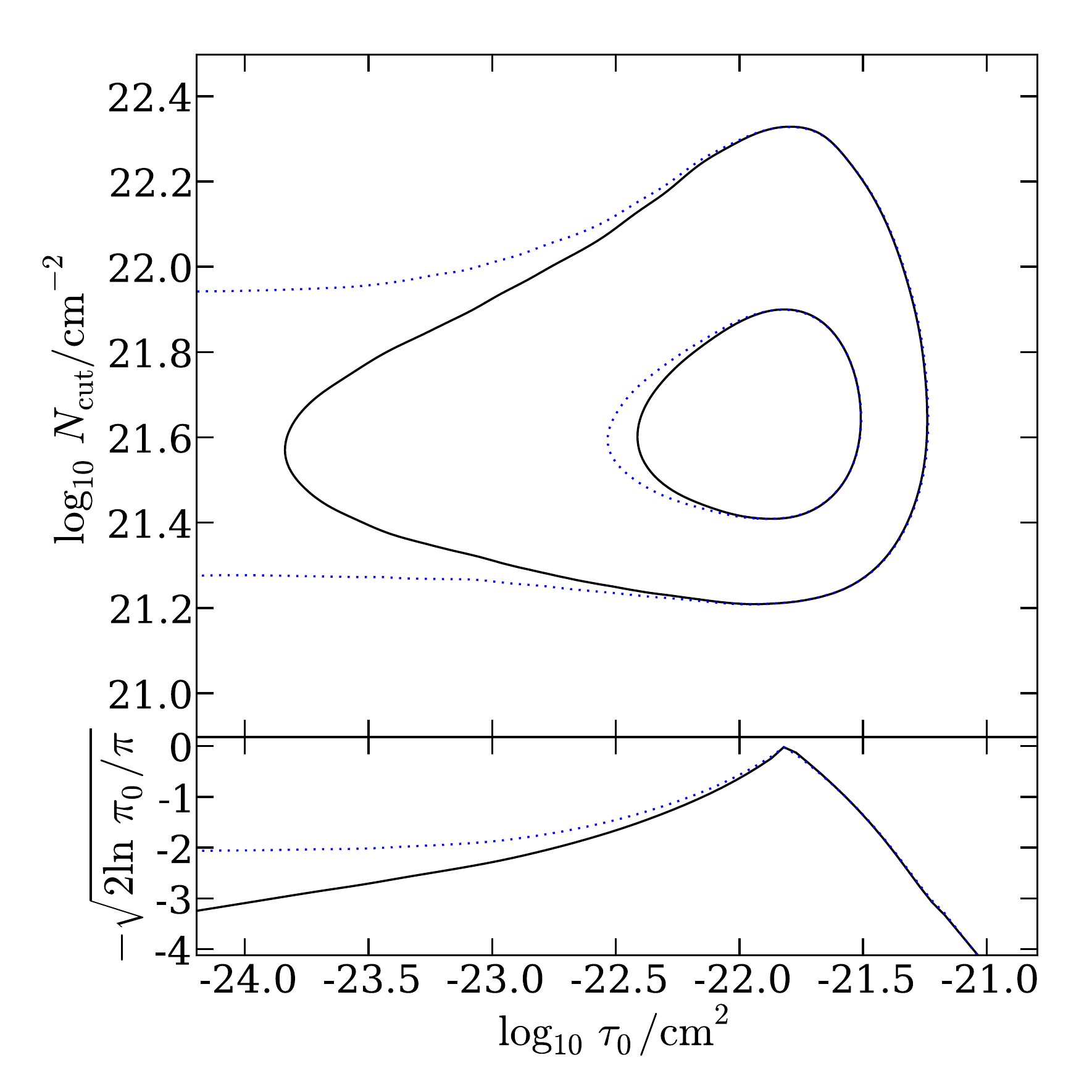}
\caption{Given the priors on $\ncut$ and $\tau_0$ listed in Table
  \ref{tab:priors}, the $1$ and $2\sigma$ contours for the
  marginalized Bayesian model fitting problem are shown (solid
  contours, top panel). These show that our model best fits the
  observed data when both an intrinsic drop in the number of high
  column density DLAs and a dust bias are in effect. To illustrate the
  effect of our prior on $\tau_0$, the dotted contours are lines of
  constant probability density for a flat log prior on $\tau_0$. The
  lower panel shows the distribution marginalized over $\ncut$ to give
  posteriors on $\tau_0$.}\label{fig:main-likelihoods}
\end{figure}

The result of our analysis is shown in the $(\tau_0,\ncut)$ plane
(marginalized over all other parameters) by the solid contours in the
upper panel of Figure \ref{fig:main-likelihoods}.  These contain
$68\%$ and $95\%$ of the total probability, corresponding to $1$ and
$2\sigma$ limits respectively. We have shown, in the lower panel, the
results of additionally integrating over $\ncut$ to gain a completely
marginalized distribution for $\tau_0$. This is plotted as $-\sqrt{2
  \ln \pi_0/\pi}$ where $\pi$ is the posterior probability density in
$\log_{10} \tau_0$ and $\pi_0 = \max(\pi)$ is an arbitrary
normalization scale. (For a normal distribution, the plotted values
$-1,-2,\dots$ would thus correspond to $1\sigma, 2\sigma, \dots$
limits.) The peak in this quantity shows that the posterior
distribution strongly suggests dust absorption with the favoured value
of $\log_{10} \tau_0 \simeq -21.8$; this is very close to the value
estimated earlier for the SMC normalization showing that the model
produces results in close accordance with expectations.

However, having used our estimate to place the {\it prior} on
$\tau_0$, it is legitimate to be concerned that our results simply
reflect this prior and hence that the data are not actually
constraining the problem.  To demonstrate that this is not the case we
have also plotted results from assuming a constant log prior on
$\tau_0$ (dotted lines in Figure \ref{fig:main-likelihoods}). The main
peak remains -- i.e. it is driven by the likelihood -- showing that
our prior, as expected, simply cuts off the otherwise infinite
distribution as $\tau_0 \to 0$ (the dotted posterior can be seen to
attain a constant value in the bottom panel, and the $2\sigma$
contours do not close in the top panel). This confirms the satisfying
result that dust obscuration of the strength implied by SMC
observations is favoured independently when analysed with our model.
We emphasize, however, that the flat prior (dotted contour) results
cannot be used in our final assessment for reasons described in
Section \ref{sec:priors}. (The dotted contours do not contain a finite
probability, but are chosen to correspond to the same probability
densities as their solid counterparts.)

\subsection{Limits on Optical Completeness}

\confitab

\begin{figure}
\includegraphics[width=0.48\textwidth]{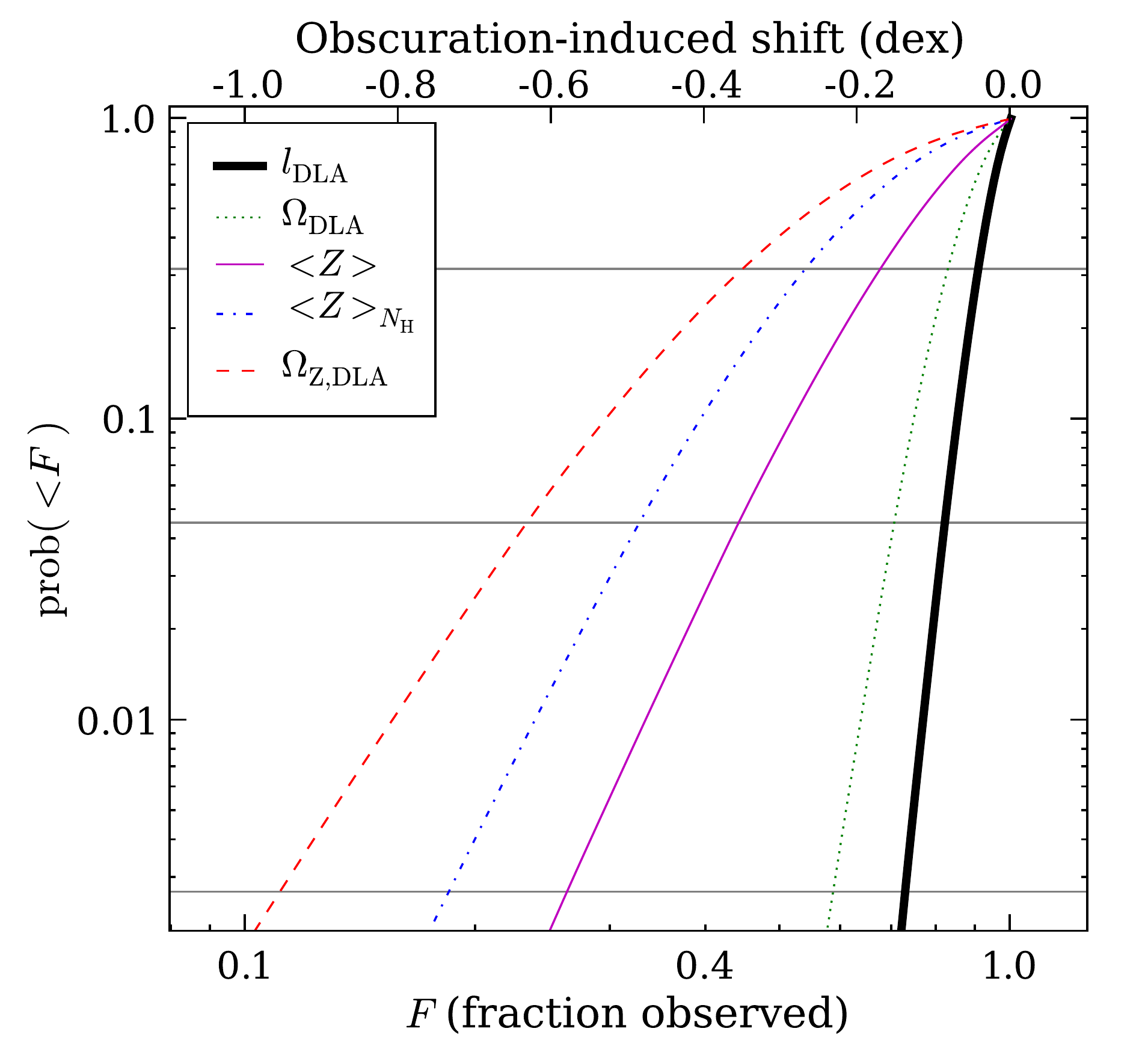}
\caption{The posterior probability for an optical fractional completeness of at
  most $F$ is plotted against $F$ for various quantities described in
  the text (see equations
  \ref{eq:l-dla}~--~\ref{eq:z-nhi-weighted}). $1$, $2$ and $3\sigma$
  completeness lower limits are given by the intersection of these
  curves with the grey horizontal lines from top to bottom
  respectively. Our results show that we are unlikely to miss a
  substantial number of DLA systems ($l_{\mathrm{DLA}}$ is almost
  complete); however metallicity-weighted measures can be more
  substantially underestimated (although not by orders of magnitude as
  previously claimed). See also Table~\ref{tab:completeness}.}\label{fig:completeness}
\end{figure}

 The most important consequence of a dust obscuration scenario is that
 various cosmological measurements may be biased. In the following, we
 will express quantities as functionally dependent on the unnormalized
 distribution function $\phi$, where $\phi=f_{\mathrm{DLA}}$ for a
 radio selected survey or $\phi=n_{\mathrm{DLA}}$ for an optically
 selected survey. We are particularly interested in the overall
 incidence rate of DLAs,
\begin{equation}
l_{\mathrm{DLA}}[\phi] \propto \int \phi(\nhi,Z)\, \dd \nhi \, \dd Z \equiv \phi_0 \label{eq:l-dla}
\end{equation}
noting that $\phi_0=f_0$ for $\phi = f_{\mathrm{DLA}}$ and
$\phi_0=n_0$ for $\phi = n_{\mathrm{DLA}}$; see equations
(\ref{eq:n0}) and (\ref{eq:f0})\footnote{Equation (\ref{eq:l-dla})
  assumes that DLAs are lost from optical surveys in direct proportion
  to $1-p_{\mathrm{detect}}$, ignoring the second-order effect from
  the reduction in the number of observed quasars. A full calculation
  shows that errors introduced by neglecting this term are at the
  percent level.}. Also of interest is the total mass density of
neutral hydrogen in DLAs,
\begin{equation}
\Omega_{\mathrm{DLA}}[\phi] \propto \int \nhi \phi(\nhi,Z)\, \dd \nhi\, \dd Z \textrm{;}
\end{equation}
the mean metallicity of DLAs,
\begin{equation}
\langle Z \rangle[\phi]  \propto \int Z \phi(\nhi,Z) \,\dd \nhi\, \dd Z / \phi_0 \textrm{;}
\end{equation}
the total mass density of metals in DLAs,
\begin{equation}
\Omega_{\mathrm{Z,DLA}}[\phi] \propto \int Z \nhi \phi(\nhi,Z) \, \dd \nhi\, \dd Z \label{eq:OZ-DLA}
\end{equation}
and the column-density weighted mean metallicity of DLAs,
\begin{equation}
\langle Z \rangle_{\nhi}[\phi] = \frac{\Omega_{\mathrm{Z,DLA}}[\phi]}{\Omega_{\mathrm{DLA}}[\phi]} \textrm{ .}\label{eq:z-nhi-weighted}
\end{equation}
Note that the missing constants of proportionality in
equations~(\ref{eq:l-dla} -- \ref{eq:OZ-DLA}) do not depend on
$\phi$. For more on the physical significance of these definitions,
see \cite{2005ARA&A..43..861W}.

The fractional completeness $F$ of any of these measurements $M$ as
measured by an optical survey is defined as
\begin{equation}
F(M) = \frac{M[n_{\mathrm{DLA}}]}{M[f_{\mathrm{DLA}}]}\label{eq:completeness}
\end{equation}
where $n_{\mathrm{DLA}}$ and $f_{\mathrm{DLA}}$ are the optical and
intrinsic distributions, defined by
equations~(\ref{eq:optical-distrib}) and~(\ref{eq:intrinsic-distrib})
respectively. $F(M)$ depends on our parameters
$\vec{r}=\left\{\alpha,\ncut,\tau_0,Z_0,\sigma_Z\right\}$; the
probability distribution for $Q=F(M)$ is written
\begin{eqnarray}
p(<Q_0) & = & \int \dd^5 \vec{r}\, p(\vec{r}) \,\theta (Q_0 - Q({\vec{r}})) \nonumber \\
& \to & \sum_{i=1}^N \theta(Q_0 - Q_i) / N  \label{eq:prob-eval}
\end{eqnarray}
where $\theta$ is the Heaviside step function, $p(\vec{r})$ is the
posterior probability density, $Q$ is any quantity dependent on the
parameters $\vec{r}$ and $Q_0$ is a value for which the cumulative
probability $p(<Q_0)$ is being calculated; when evaluating from the
MCMC chain, $i$ ranges over the models in the chain and $N$ is the
number of steps.  We will also be interested in the expected value of
$Q$,
\begin{equation}
E(Q) = \int \dd^5 \vec{r}\, p(\vec{r}) Q(\vec{r}) \to \sum_{i=1}^N Q_i / N \textrm{.}
\end{equation}
The results for the five quantities defined in equations
(\ref{eq:l-dla}~--~\ref{eq:z-nhi-weighted}) are shown in Figure
\ref{fig:completeness} and Table \ref{tab:completeness}. For each
quantity $M$, the plot shows the cumulative probability
$p(\textrm{completeness}<F)$ and the table specifies the expected
value $E(F)$ along with confidence intervals at $1,2,3 \sigma$
(i.e. $(F_0,F_1)$ such that $p(F_0 <
\textrm{completeness}<F_1)=67\%,95\%$ and $99.7\%$). The immediate
conclusion is that optical samples are likely to be biased but at a
level significantly smaller than many previous studies have
claimed. Simple quantities such as the overall incidence rates of DLAs
($\ldla$) are likely to be almost unaffected (with only a $7\%$
expected underestimate and $<10\%, 19\%$ at $1, 2\, \sigma$
confidence). On the other hand, quantities which are weighted towards
higher column densities of metals suffer more from the effects of
obscuration. The total DLA mass in \hi~($\Omega_{\mathrm{DLA}}$) is
unlikely to have been underestimated by more than $30 \%$ ($2\sigma$
limit), but a heavily weighted quantity such as the total mass of {\it
  metals} in DLAs ($\Omega_{Z,\mathrm{DLA}}$) is underestimated by
about a factor of two, or at most $78 \%$ ($2 \sigma$). Note that this
nonetheless results in a relatively modest worst-case shift in the
mean metallicity of $<0.4\, \dex$ ($2 \sigma$ limit) or $<0.5\, \dex$
for the column density weighted metallicity ($2 \sigma$ limit).

In some cases, we can compare the completeness limits derived above
with analogous estimates by other authors.  \cite{2006ApJ...651...51T}
compared the column density distributions of SDSS DR3 and
\textsc{Corals} radio samples, concluding that optical determinations
of $\Omega_{\mathrm{DLA}}$ underestimated the true value by around
15\%, very close to our own expected value of 13\%. This is perhaps
unsurprising since the information used in this earlier work is a
subset of our own dataset.  Estimates which take into account the
optically determined metallicity distribution (but not the comparison
with radio-selected quasars) are to be found in
\cite{2005A&A...444..461V}.  The authors give values for the
completeness of $l_{\mathrm{DLA}}$ of $50 - 70\%$ and claim
$\Omega_{\mathrm{DLA}}$ is underestimated by at least $50\%$ (their
section 6.5).  These estimates are inconsistent with our $3 \sigma$
limits for the minimum completeness, and differ substantially from our
expected value of $E\left[F(\Omega_{\mathrm{DLA}})\right] = 87
\%$. Further, although \cite{2005A&A...444..461V} did not give
completeness statistics comparable to ours for their metallicity
distributions, they suggest metallicities are underestimated by
factors of 5 to 6, a shift of about $0.8\,\dex$, again incompatible
with our $3\sigma$ limits. We explore possible explanations for these
differences in Section~\ref{sec:conclusions}.

\subsection{Expected Dust Reddening}

Because the optical depth of dust rises rapidly towards shorter
wavelengths, observed quasars obscured by dust are expected to exhibit
statistically redder spectra than their unobscured counterparts.  This
effect is discussed in the introduction, but we did not use the
results of recent dust reddening studies
\citep{2008A&A...478..701V,2004MNRAS.354L..31M} as priors in our
model, since the uncertainties of these authors' analyses are quite
different in nature from the uncertainties in our model.  However, we
should check that our results are indeed compatible with the observed
reddening effect.

We caution that our estimate will assume a proportionality between the
colour shift $E_{B-V}$ in the DLA rest-frame and the strength of the
overall obscuration, calculated according to the SMC extinction law
measured at $1900\,\angst$. This assumption is not fully justified
given the differing DLA redshifts over the sample although, as before,
we expect that performing the calculation assuming mean values in this
way should not introduce a substantial bias.

The expected reddening effect of a DLA on the background quasar is
\begin{eqnarray}
\langle E_{B-V}\rangle[\phi] \hspace{-0.2cm} & = & \hspace{-0.2cm} \left(\frac{E_{B-V}}{\tau(1900\angst)}\right)_{\mathrm{SMC}}\,\tau_0  \nonumber \\*
&&\hspace{0.1cm}\times \int \dd \nhi\, \dd Z  \frac{\nhi \ffe(Z)Z \phi(\nhi,Z)}{\ffe(Z_0)Z_{0} \phi_0}
\end{eqnarray}
where the first factor is evaluated from the SMC reddening curve
giving $E_{B-V}/\tau(1900\angst) \simeq 0.12$, the normalizing
constant $\phi_0$ is defined as usual (equation~\ref{eq:l-dla}) and
$\phi = f_{\mathrm{DLA}}$ for a radio survey or
$\phi=n_{\mathrm{DLA}}$ for an optical survey.

\begin{figure}
\includegraphics[width=0.5\textwidth]{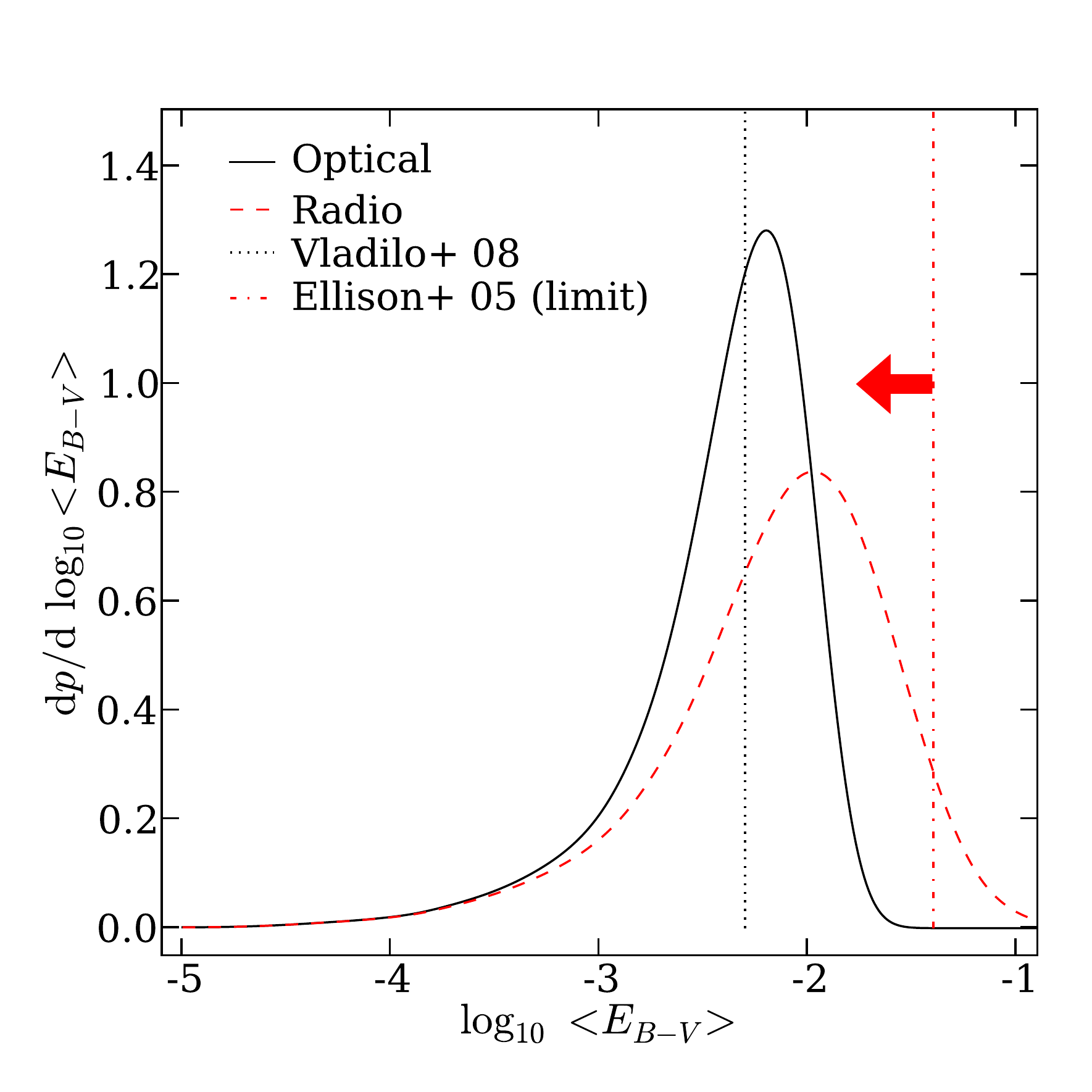}
\caption{Dust reddening in our models. The solid curve gives the
  probability density, according to our posterior distribution, of
  various mean reddenings expected in an optical sample (comparing
  quasar spectra to their appearance in the absence of a DLA). The
  dashed curve gives the same statistics for a radio sample (no dust
  bias). The vertical dotted line gives the measured reddening in SDSS
  DR5 \protect\citep{2008A&A...478..701V}, while the dash-dotted line
  gives the direct upper limit on radio-selected quasar reddening
  derived by \protect\cite{2005AJ....130.1345E}. We emphasize that our model
  fitting is performed without using any such constraints, so that the
  agreement of both statistics is an independent validation of our
  results.}\label{fig:dust-red}
\end{figure}

The posterior distribution is evaluated according to
equation~(\ref{eq:prob-eval}) setting $Q=\langle E_{B-V} \rangle$; the
results are shown in Figure \ref{fig:dust-red} with confidence
intervals listed in Table \ref{tab:completeness}. It is satisfying
that our $1\sigma$ interval $-2.6< \log_{10} \langle E_{B-V} \rangle <
-2.0$ for an optical survey agrees with the result $\log_{10} \langle
E_{B-V} \rangle \simeq -2.2\pm 0.1$ of \cite{2008A&A...478..701V}
\cite[and is consistent with the upper limit
  of][]{2004MNRAS.354L..31M}. The expected reddening effect of DLAs in
radio samples (dashed line in Figure \ref{fig:dust-red}) is more
pronounced than that in optical samples (solid curve), since the
average radio-selected DLA will have a higher column density of metals
(no dust bias). This is, however, compatible with the limit on
radio-selected DLA reddening $\langle E_{B-V} \rangle<0.04$ from
\cite{2005AJ....130.1345E}.

\subsection{Internal Consistency and Driving Factors}\label{sec:intern-cons-driv}

Given that previous studies of the $(\nhi, Z)$ evidence for dust
obscuration have generally pointed to more pronounced effects than
indicated by radio-selected surveys (see 
Introduction), we should check that this tension is not present in our
analysis; if so this could point to a deficiency in the model,
limiting the usefulness of the results. A severe tension, with one
dataset requiring different parameters from others, would result in
the posterior parameters being pushed to intermediate values
incompatible with estimates from individual likelihood terms in
equation~(\ref{eq:tot-likeli}).

The optical completeness in our final analysis is expected to be $\sim
90\% - 100\%$ (Table \ref{tab:completeness}), giving an expected
number of radio-selected DLAs $11.8 \lsim \lambda \lsim
13.1$. Comparing to the actual number, $k=17$, shows that the radio
observations actually detect a slightly larger number than our model
has predicted -- in other words, they prefer {\it stronger} dust
absorption.  However, because of the small path length of existing
radio surveys, the Poisson likelihood (\ref{eq:linedens-likeli}) has a
wide variance $\sigma = \sqrt{\lambda} \sim 3.5$. The consequence of
this is that the overall $1\sigma$ region is almost entirely contained
within the $1\sigma$ region for the line density data. This shows
there is not a substantive tension between these datasets in our
analysis. Because of this very consistency (coupled with the wide
variance) excluding the line density likelihood from the final
analysis makes only minor differences to the results (a fact we
explicitly verified); however we have retained it for completeness.

It is worth briefly investigating which data are most powerful in
producing constraints on dust effects, especially bearing in mind our
comments in the Introduction that the apparent anticorrelation of
$\nhi$ and $Z$ in optically selected DLA samples can be reproduced
without any dust effects whatsoever (see Figure \ref{fig:models}).
Concretely, our optical sample of zinc and iron metallicities (Section
\ref{sec:likelihood}) is calculated to have a Spearman rank
correlation statistic $r=0.055$, giving a two-tailed p-value of $0.55$
(i.e. a sample of the same size with random uncorrelated values will
show the same or greater levels of apparent correlation in more than
$55\%$ of cases). This leads to the expectation that, on its own, the
optical data can place only an upper limit on the effect of dust
obscuration.  We explicitly verified that this is the case by running
our analysis without any radio constraints.

In fact, the major factor in determining our results is the comparison
of radio-observed and optically-observed distributions of column
densities and metallicities. These lead to the positive {\it
  detection} of dust obscuration effects even with neutral priors that
allow for no dust in DLAs whatsoever (dotted blue lines in Figure
\ref{fig:main-likelihoods}; see Section \ref{sec:results}). Because
this detection so closely matches estimates calculated from
observations of the interstellar medium in the SMC, we may have some
confidence that our final results are meaningful.

\section{Discussion and Conclusions}\label{sec:conclusions}

In this work, we have analysed radio- and optically-selected DLA
samples to produce an overall picture of dust obscuration. We first
noted that the distribution of optically selected DLAs in the
$(\nhi,Z)$ plane does not point unambiguously to significant dust
obscuration. In fact, it is quite possible to form reasonable models
in which high metallicity, high \hi~column density DLAs are rarely
seen simply because the product of the metallicity and column density
distributions is small in this region (Figure \ref{fig:models}, left
panel).

We assembled a simple model of DLA dust obscuration in which the
intrinsic DLA distribution is separable in the $(\nhi,Z)$ plane and a
tuneable dust parameter $\tau_0$ obscures a variable fraction of DLAs
from optical samples based on the total column density in dust
(modelled as $\nhi Z \ffe(Z)$). We then assessed this model
using a Bayesian parameter estimation approach with a likelihood based
on four sets of observational data: an optically-selected sample of
column densities and metallicities \cite[based on
][]{DZinprep}; an equivalent radio-selected survey
\citep[\textsc{Corals}; ][]{2001A&A...379..393E,2005A&A...440..499A};
the SDSS statistics for observed column densities of DLAs
\citep{2005ApJ...635..123P}; and a comparison of the incidence rates
of DLAs towards optical- and radio-selected quasars
\citep{2001A&A...379..393E}.

Table \ref{tab:completeness} summarises the observational predictions
of our model.  The results do not allow a large hidden fraction of
DLAs; thus simple quantities such as their line density
($l_{\mathrm{DLA}}$) are not significantly underestimated in existing
surveys.  Quantities weighted towards higher metal column densities
are (as expected) less well constrained by optical surveys. However,
there is still relatively little room for manoeuvre; in particular,
our statistics for the metallicity suggest that substantial
(i.e. $\gsim 1\,\dex$) dust-induced corrections of the type sometimes
invoked to reconcile models with data
\citep[e.g.][]{2003ApJ...598..741C,2004MNRAS.348..435N} are not
supported by the data.

Our model is similar to that of \citeauthor{2005A&A...444..461V}
(2005; henceforth VP05) but we arrive at some qualitatively different
conclusions. This may be due to the expanded sample now available, but
it is worth noting that our analysis also differs in many details:

\begin{enumerate}

\item We have used a Bayesian approach, being careful to avoid
  focussing on our peak likelihood model but rather analysing the
  entire posterior distribution. This leads to well-defined
  statistical limits on the effects under consideration.
\item We have used a substantially larger optical sample of DLAs and
  additionally considered radio-selected and SDSS observations. We
  found that on their own optical samples are rather poor at
  constraining the effects of dust (Section
  \ref{sec:intern-cons-driv}) whereas adding radio samples returns
  results which are promisingly consistent with estimates from
  sightlines through the SMC.
\item We formulate the likelihood for each observation rather than
  simply fitting the distribution using a $\chi^2$ minimization
  technique. For instance, one may not assume that the statistics of
  high resolution optical samples trace the underlying $\nhi$
  distribution since observers choose their targets using a variety of
  criteria. Such an assumption leads VP05 to estimate a shallower
  distribution of $\nhi$ values than is revealed by the SDSS (powerlaw
  indices $-1.5$ and $-1.8$ respectively), for example. This could
  plausibly bias the estimation of dust effects.
\item We have used a lognormal distribution for metallicity which we
  argue eliminates some further biases (equation
  \ref{eq:metal-lognormal} and discussion thereafter, or for more
  details see Appendix \ref{sec:choice-fit-funct}). Also, we have used
  iron abundances where zinc are unavailable, since this traces the
  low metallicity end of the distribution. While iron is refractory and
  therefore can underestimate true metallicities, this is a small
  effect at low metallicities where zinc becomes systematically hard
  to detect (Figure \ref{fig:vp05-depletion}). We believe this
  situation is preferable to ignoring the low metallicity tail of the
  distribution.
\item 
  For the column density distribution, we allowed for an exponential
  cut-off at $\nhi\gsim \ncut$ (equation \ref{eq:colden-schechter}).
  If this parameter were unnecessary, our posteriors would have
  automatically pushed $\ncut$ to high values -- but this was not the
  case (Figure \ref{fig:main-likelihoods}). (The likelihood is also
  sufficiently peaked that only an extreme prior would reverse this
  trend.) Removing the intrinsic exponential cut-off in $\nhi$ would
  have at least two problematic effects.  Firstly, it forces a
  substantial increase in the deduced effects of dust obscuration,
  since these alone must account for the drop in observed high $\nhi$
  absorbers (as illustrated by the central panel of Figure
  \ref{fig:models}). Secondly, it causes estimates for completeness of
  weighted quantities such as $\Omega_{\mathrm{DLA}}$ to converge
  extremely slowly, with a substantial contribution arising from
  extremely high column densities. VP05 impose arbitrary cut-offs at
  high $\nhi$ to estimate such effects but the results are sensitive
  to the cut-off chosen.
\item We have used a somewhat simplified obscuration model, arguing
  that fine details are absorbed into our parameter definitions and do
  not affect estimates for quantities of observational interest.
\end{enumerate}

There are two notable omissions in our modelling. Firstly, we assumed
that the intrinsic cutoff $\ncut$ was not dependent on metallicity.
But taking seriously the suggestion of \cite{2001ApJ...562L..95S} that
the physical mechanism for preventing arbitrarily high $\nhi$
absorbers is the conversion of \hi~into H$_2$, one would expect the
characteristic transition column density to be linked to the presence
of dust (an essential catalyst in the efficient production of
H$_2$). In fact, this would give a neat explanation for the
coincidence of intrinsic and dust-induced cut-offs (by which we mean
$\ncut \simeq \tau_0^{-1}$). But if this effect depends on
metallicity, as is plausible, it should introduce intrinsic
correlations in the ($\nhi$,$Z$) plane; these would be in the same
sense as dust obscuration effects.  Since our likelihood is largely
controlled by the comparison of radio and optical data (Section
\ref{sec:intern-cons-driv}) which would not essentially be changed in
such a scenario, it is likely that our analysis is robust. Nonetheless
without a more specific physical model it is hard to assess this in
more detail.

Secondly, we have not included a model of gravitational lensing by the
host halos of DLAs. There is some evidence in the SDSS sample of a
correlation between $\nhi$ and the background quasar luminosity which
can be explained by this effect
\citep{2004MNRAS.354L..31M,2005ApJ...635..123P}. Our simulations
\citep{PontzenDLA} suggest that, in fact, the metallicity $Z$ of a
system is a better indication of its mass than $\nhi$. Thus any
lensing effect will presumably be correlated with metallicity.  If so,
the resulting entanglement may cause us to underestimate the dust bias
-- although if the processes genuinely compensate each other, the
completeness limits are unchanged! (Gravitational lensing being
monochromatic, this could only work in one waveband.) A full
assessment of this possibility awaits future work.

It seems likely that DLA dust biasing is a real but minor effect; all
observational constraints are essentially consistent with this
conclusion. The fractional completeness of optically-determined values
for observable quantities depend on their weighting towards higher
metal column densities. The least affected quantity is the overall
incidence rate $\ldla$ which is expected to be $93\%$ complete; the
most affected quantity is the mass of metals in DLAs
$\Omega_{\mathrm{Z,DLA}}$ which is nonetheless expected to be
underestimated only by a factor of about two.

\section*{Acknowledgments}

We thank Giovanni Vladilo, Celine Peroux, Sara Ellison and the
referee, Mike Edmunds, for many helpful comments which improved the
quality of this paper.  Miroslava Dessauges-Zavadsky, Sara Ellison and
Michael Murphy kindly made their compilation of DLA abundances
available to us prior to publication; we would also like to
acknowledge Chris Akerman, Varsha Kulkarni, Jason Prochaska and their
collaborators for making compilations of metallicity measurements in
DLAs generally available. AP is supported by a STFC studentship and
scholarship at St John's College, Cambridge and gratefully
acknowledges several helpful conversations with Steve Gratton, Natasha
Maddox and Antony Lewis.

\bibliographystyle{mn2e} {\small \bibliography{../../refs}}

\appendix

\section{Choice of Fitting Functions}\label{sec:choice-fit-funct}
\begin{figure*}
\includegraphics[width=0.5\textwidth]{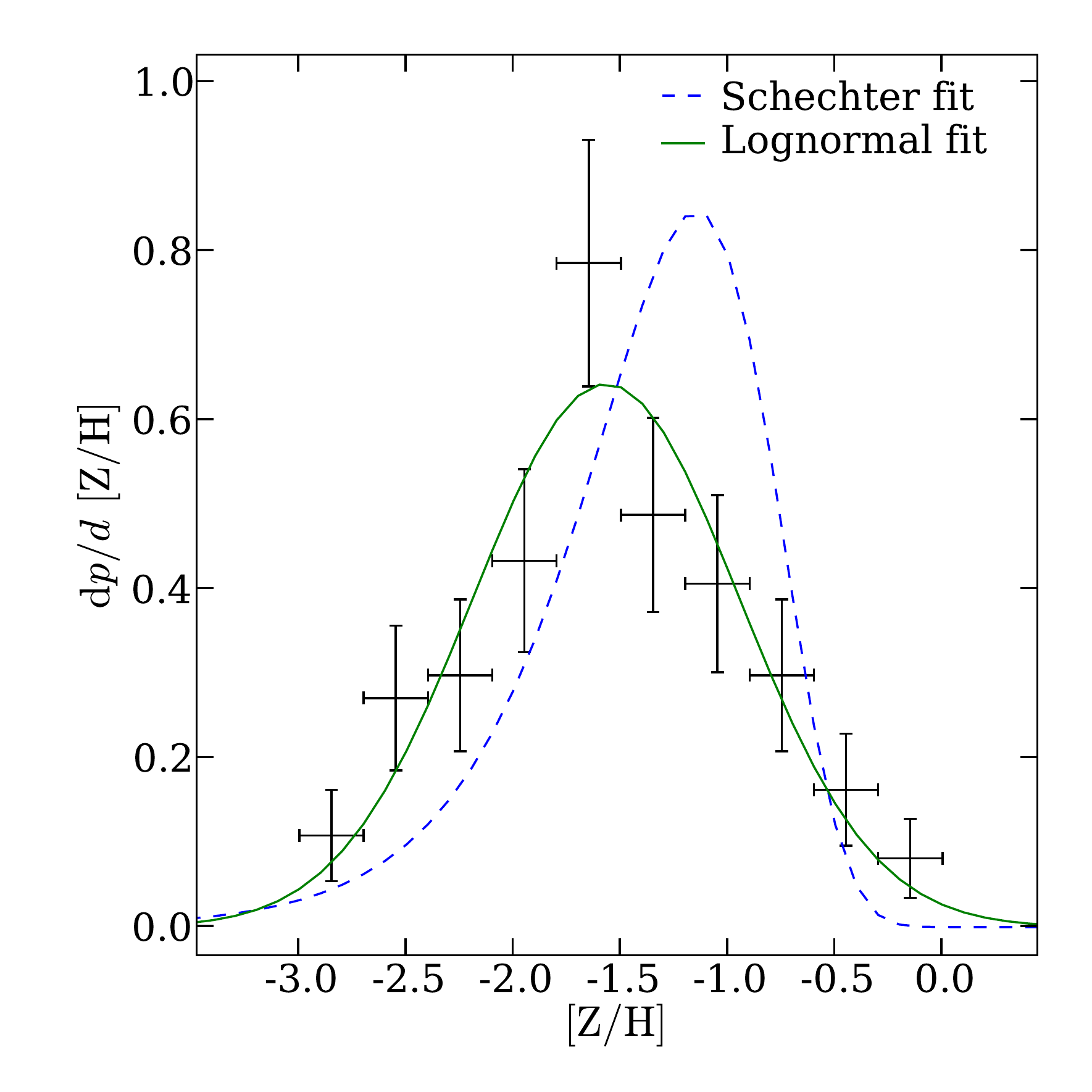}\includegraphics[width=0.5\textwidth]{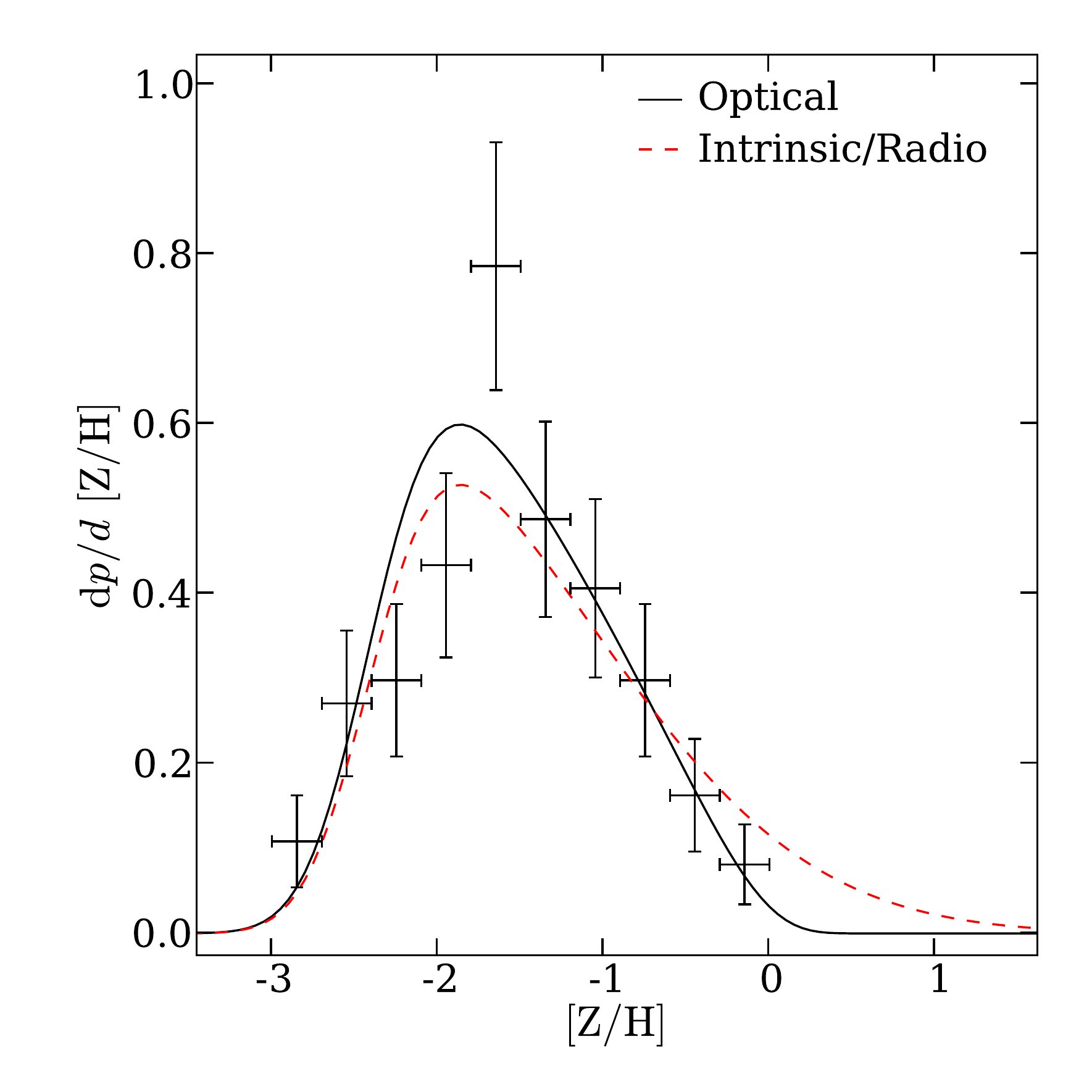}
\caption{The points with error bars show our optical sample of
  metallicities based on \protect\cite{DZinprep} and described in
  Section \ref{sec:likelihood}. (Note that the binning is for
  illustrative purposes only and is not part of the analysis). In the
  left panel the solid and dashed lines show simple best fit lognormal
  and Schechter distributions respectively. The Schechter fit to the
  observed optical distribution is strongly disfavoured (see text for
  details) and therefore employing this function for the intrinsic
  distribution may artificially disfavour small bias scenarios. In the
  right panel, we illustrate a model in which the underlying
  metallicity distribution $f_Z$ (shown by the dash line) is strongly
  skewed in log space, but dust absorption hides the long tail to high
  metallicities in optically selected surveys (solid line). In the
  illustrated model, the skewness parameter $\zeta$ is $5.8$ and the
  dust obscuration ($\tau_0 =-22.0$) hides the tail almost
  completely. This model should be discounted by a prior on allowed
  metallicities -- even if the radio sample of DLAs is not strong
  enough to rule it out, the model includes significant numbers of
  DLAs with $Z\gg 5 Z_{\odot}$, greater than the values measured in
  even the most massive galaxies \protect\citep{2005ApJ...621..673T}.}\label{fig:schechter-rubbish}
\end{figure*}

The choice of Schechter and lognormal distribution
functions (equations~\ref{eq:colden-schechter}
and~\ref{eq:metal-lognormal}) for our intrinsic column density and
metallicity distributions respectively is somewhat arbitrary and we
should ensure our choice is as fair as possible given our prior
knowledge. 

The column density distribution choice is relatively easy to
justify. We know that the obscured distribution as uncovered by the
SDSS statistics is well fitted by a Schechter function
\citep{2005ApJ...635..123P}, which consists of a power-law suppressed
at high column densities by an exponential decline. This decline can
arise due to an intrinsic cut-off ($\ncut$) or due to the exponential
dust suppression term in equation (\ref{eq:optical-distrib}). We
verified that our choice of column density distribution function does
not bias our results by using only the SDSS likelihood
(equation~\ref{eq:sdss-likeli}) to produce a posterior prediction for
$\tau_0$ which simply returned our prior.

The form of the metallicity distribution presents more serious
difficulties. Because of the small-number statistics in the radio
samples, we know most about the optically determined (obscured)
distribution, which reads
\begin{equation}
n_Z(Z) = \int \dd \nhi \, f_{Z}(Z) f_{N}(\nhi) \, p_{\mathrm{detect}}\left[\tau(\nhi,Z)\right] \textrm{ .}
\label{eq:n_Z}
\end{equation}
It should be clear that the intrinsic distribution $f_{Z}(Z)$ is only
recoverable from the data once we know the strength of the dust
absorption.  Even then, with finite statistics one can never rule out
the existence of a population of very high metallicity absorbers which
are hidden from view. We therefore need to make an ad hoc
parameterization of $f_Z$ which encapsulates our prejudice that (i) the
distribution function should change smoothly and (ii) the distribution
function is likely to be unimodal. We will not discuss any models
which fail to satisfy these conditions, but accept they could change
results substantially.

For our main results, we chose to use the lognormal distribution.
\cite{2005A&A...444..461V} contended that a Schechter function
provides a more generic fit, arguing that the shape of the high- and
low-metallicity tails can be independently controlled. However, since
both lognormal and Schechter fits have only two parameters, this claim
should be interpreted cautiously. Given any two-parameter fit, once
the mean and variance are specified the exact distribution function
and hence its higher moments (such as the skewness) are fixed. Thus we
should investigate which distribution function better encapsulates our
knowledge of the systems; if necessary, extra parameters can then be
introduced to compensate for deficiencies. The lognormal distribution
is a fairly generic choice; further it is supported as a choice for
$f_Z$ by simulations \citep{PontzenDLA} although it is hard to know
how much weight to assign to such support.

With our current data we find that the Schechter function provides a
very poor representation of the obscured distribution $n_Z(Z)$. Figure
\ref{fig:schechter-rubbish} (left panel) shows the best fit lognormal
and Schechter distributions; with flat priors on the expectation and
variance, the latter distribution is disfavoured in log evidence by
more than 10, i.e. the probability of the data arising given the
latter distribution is more than 20,000 times smaller. Employing this
function for the intrinsic distribution $f_Z(Z)$ is therefore likely
to bias results against any scenario in which $n_Z \simeq f_Z$,
i.e. where dust obscuration is small.

However, accepting that the lognormal distribution may be too
restrictive a form for $f_Z$ (even if it fits $n_Z$ well) we
investigated the effect of generalising the metallicity distribution
to a three-parameter family of distributions which allow for skewing
the underlying $f_Z$. For this purpose, we have used a log skew-normal
distribution. The skew-normal distribution \cite[see][and references
  therein]{AzzaliniSkewNormal} is written
\begin{equation}
\xi(Z;\zeta,M,S) = 2 \psi\left((Z-M)/S\right) \, \Psi\left(\zeta(Z-M)/S\right)
\end{equation}
where $\psi$ and $\Psi$ are respectively the probability density and
cumulative probability of the normal distribution. It is remarkable,
but simple to show, that this distribution is normalized for all
values of $\zeta$. For $\zeta=0$, the distribution is exactly normal;
as $\zeta\to +\infty, -\infty$ one obtains the half-normal
distribution for $Z>M$ and $Z<M$ respectively.  In between these
extremes, $\zeta$ smoothly interpolates between models of varying
skewness.

When $\zeta$ is allowed to take any value it is possible to find
models with large tails of high metallicity DLAs in which dust
obscuration makes the optical distribution compatible with the
data. An extreme case is illustrated in the right panel of Figure
\ref{fig:schechter-rubbish}; the dashed line shows the intrinsic
(strongly skewed) distribution while the solid line shows the observed
(dust obscured, nearly symmetric) distribution.

The radio sample is somewhat too small to fully rule out such cases,
but we should impose a prior reflecting our knowledge of metallicities
in the Universe. In particular, it would be extremely surprising to
find a significant number of systems with $Z>10 Z_{\odot}$ \cite[see,
  e.g.,][in which the centres of $z=0$ early type galaxies are shown
  not to exceed even $3 Z_{\odot}$]{2005ApJ...621..673T}.  Therefore,
in a test run of our markov chain, we allowed $\zeta$ to vary with
uniform prior but imposed a ``brick wall'': models predicting greater
than one in 1000 intrinsic DLA systems of $Z>10 Z_{\odot}$ were given
zero prior probability. This is, of course, an arbitrary choice and
will be model-dependent in its implications. But it is a simple
first-order approximation, allowing a model with more complex
behaviour while imposing our knowledge of direct observations of
galaxies.

Comparing this choice with our main ($\zeta=0$) results, the
differences in our posterior distribution were at the percent level
and made no difference to our qualitative conclusions presented in the
main paper. As the high-metallicity wall is relaxed, allowing more
$Z>10 Z_{\odot}$ systems, the constraints are weakened; if one imposes
no such prior, allowing systems of arbitrarily high metallicity, $1
\sigma$ confidence intervals become $0.83<F(\ldla)<0.95$, $0.34<F(\langle Z
\rangle )< 0.69$ and $0.12<F(\Omega_{Z,\mathrm{DLA}})<0.50$. However, we
emphasize that much of the obscured cross-section is then in
exceptionally high metallicity DLAs with $Z > 10 Z_{\odot}$ -- such 
a model seems very unlikely.

In future, it will be possible to place tighter constraints on these
model freedoms by obtaining expanded samples of DLAs from
radio-selected QSO spectra.  Although further blind radio surveys are
relatively slow to reduce the variance of incidence rate statistics
(fractional errors for $N_{\mathrm{radio}}$ DLAs scale as
$1/\sqrt{N_{\mathrm{radio}}}$), high resolution follow-up spectroscopy
greatly increases the model-discerning power of the radio
observations. Simulations based on our peak posterior model showed
that, with an increase in sample size to $N_{\mathrm{radio}} \simeq
35$ (approximately twice the current number of \textsc{Corals} DLAs
with measured metallicities), models with a high-metallicity tail
could be independently rejected by the DLA sample. Conversely if a
significant high-metallicity skew-normal tail exists but is hidden in
optical samples, such a modestly expanded radio sample would be
sufficient to reveal its existence. We therefore encourage observers
to pursue further searches for DLAs in complete (i.e. fully optically
identified) samples of radio-loud QSOs.

\end{document}